# Title: Experimental evidence for Berry curvature multipoles in antiferromagnets

**Authors:** Soumya Sankar[1]†, Ruizi Liu[2]†, Xue-Jian Gao[1]†, Qi-Fang Li[3,4]†, Caiyun Chen[1,5]†, Cheng-Ping Zhang[1], Jiangchang Zheng[1], Yi-Hsin Lin[1], Kun Qian[2], Ruo-Peng Yu[1], Xu Zhang[4], Zi Yang Meng[4], Kam Tuen Law[1], Qiming Shao[1,2,6,7]*, Berthold Jäck[1,6]*

**Affiliations:**

[1]The Hong Kong University of Science and Technology, Department of Physics, Clear Water Bay, Kowloon, Hong Kong SAR, China

[2]The Hong Kong University of Science and Technology, Department of Electronic and Computer, Clear Water Bay, Kowloon, Hong Kong SAR, China

[3]The Institute for Solid State Physics, The University of Tokyo, Chiba, Japan

[4]Department of Physics and HKU-UCAS Joint Institute of Theoretical and Computational Physics, The University of Hong Kong, Pokfulam Road, Hong Kong SAR, China

[5]Institute for Advanced Studies, The Hong Kong University of Science and Technology, Clear Water Bay, Kowloon, Hong Kong SAR, China

[6]IAS Center for Quantum Technologies, The Hong Kong University of Science and Technology, Clear Water Bay, Kowloon, Hong Kong SAR, China

[7]Guangdong-Hong Kong-Macao Joint Laboratory for Intelligent Micro-Nano Optoelectronic Technology, The Hong Kong University of Science and Technology, Hong Kong SAR, China

†These authors contributed equally.

*Correspondence should be addressed to Berthold Jäck (bjaeck@ust.hk) and Qiming Shao (eeqshao@ust.hk).

**Abstract:** Berry curvature multipoles appearing in topological quantum materials have recently attracted much attention. Their presence can manifest in novel phenomena, such as nonlinear anomalous Hall effects (NLAHE). The notion of Berry curvature multipoles extends our understanding of Berry curvature effects on the material properties. Hence, research on this subject is of fundamental importance and may also enable future applications in energy harvesting and high-frequency technology. It was shown that a Berry curvature dipole can give rise to a $2^{nd}$ order NLAHE in materials of low crystalline symmetry. Here, we demonstrate a fundamentally new mechanism for Berry curvature multipoles in antiferromagnets that are supported by the underlying magnetic symmetries. Carrying out electric transport measurements on the kagome antiferromagnet FeSn, we observe a $3^{rd}$ order NLAHE, which appears as a transverse voltage response at the $3^{rd}$ harmonic frequency when a longitudinal a.c. current drive is applied. Interestingly, this NLAHE is strongest at and above room temperature. We combine these measurements with a scaling law analysis, a symmetry analysis, model calculations, first-principle calculations, and magnetic Monte-Carlo simulations to show that the observed NLAHE is induced by a Berry curvature quadrupole appearing in the spin-canted state of FeSn. At a practical level, our study establishes NLAHE as a sensitive probe of antiferromagnetic phase transitions in other materials, such as moiré superlattices, two-dimensional van der Waal magnets, and quantum spin liquid candidates, that remain poorly understood to date. More broadly, Berry curvature multipole effects are predicted to exist for 90 magnetic point groups. Hence, our work opens a new research area to study a variety of topological magnetic materials through nonlinear measurement protocols.

# Introduction

In materials with topologically non-trivial electronic states, a finite Berry curvature $\mathbf{\Omega}$ has a profound influence on the electrical, optical, and thermal materials properties [1,2]. Microscopically, $\mathbf{\Omega}$ is related to the topology of the quantum-mechanical wave function and can be regarded as the momentum space analogue of a magnetic flux density. Hence, a finite $\mathbf{\Omega}$ can induce an anomalous Hall effect (AHE), which appears as a finite Hall voltage $V_\mathrm{H}$ in electrical transport measurements [3]. The Berry curvature contribution to the Hall effect arises from its momentum space integration over the occupied electronic states $\mathcal{M} \propto \int f_0 \Omega_z \, d\mathbf{k}$ where $f_0$ is the Fermi-Dirac distribution. Effectively, it can be viewed as a Berry curvature monopole $\mathcal{M} \propto \rho_\mathrm{xy}^{-1}$. Because of the direct connection between $\mathbf{\Omega}$ and the Hall resistivity $\rho_\mathrm{xy}$, the AHE plays a key role in the investigation of topological quantum materials, such as Weyl [4,5] and Dirac [6] fermion systems, and the Chern insulating state in the quantized Hall limit [7–9].

It has been recognized that Berry curvature multipole moments, such as dipoles $\mathcal{D}_{\alpha\beta}$ [10,11], quadrupoles $\mathcal{Q}_{\alpha\beta\gamma}$, hexapoles, and octupoles [12] can generally exist. These Berry curvature multipoles derive from the momentum space derivatives of $\mathbf{\Omega}$, e.g., $\mathcal{D}_{\alpha\beta} \propto \int f_0 \partial_\alpha \Omega_\beta d\mathbf{k}$ and $\mathcal{Q}_{\alpha\beta\gamma} \propto \int f_0 \partial_\alpha \partial_\beta \Omega_\gamma d\mathbf{k}$ where $\alpha, \beta, \gamma$ denote the spatial directions. Finite Berry curvature multipole moments modify the material properties, and their presence can be experimentally detected, for example, through measurements of the nonlinear anomalous Hall effect (NLAHE). Here, multipole moments induce a finite Hall voltage $V_\mathrm{H}^{n\omega}$ at the higher harmonics $n\omega$ ($\omega = 2, 3, ...$) of a longitudinally applied a.c. current with frequency $\omega$. A finite $\mathcal{M}$ permits a first order response $V_\mathrm{H}^{1\omega}$ when time-reversal symmetry $\mathcal{T}$ is broken, the conventional linear anomalous Hall effect. The breaking of crystalline inversion symmetry $\mathcal{P}$ [10] can yield a finite $\mathcal{D}_{\alpha\beta}$, which permits a second order Hall response $V_H^{2\omega}$ even when $\mathcal{T}$ is preserved. This 2nd order nonlinear Hall effect was initially observed in bilayer WTe$_2$ [13], which is non-magnetic. Interestingly, it has been predicted that the magnetic point group symmetries of antiferromagnetic materials can support even higher order Berry curvature multipole moments, such as quadrupole, hexapole, and octupole moments, which should manifest in NLAHEs up to the 5th order. Crucially, the origin of these higher order multipole moments is qualitatively different from that of the previously observed Berry curvature dipole moment [10,13–15], which only requires a low crystalline symmetry. To date, experimental evidence for higher-order Berry curvature effects in antiferromagnets is missing.

Here, we report the observation of a Berry curvature quadrupole (BCQ) induced 3rd order NLAHE in electric transport measurements conducted on the kagome antiferromagnet (AFM) FeSn [16,17]. The observation of this NLAHE is consistent with expectations from our symmetry analysis, as well as tight-binding and *ab-initio* calculations that predict the presence of a BCQ in the spin-canted state of FeSn. Interestingly, the observed NLAHE signal is strongest above room temperature and sensitive to the symmetry of the magnetic order parameter. Our scaling analysis in terms of the charge carrier scattering time reveals the intrinsic BCQ origin of the 3rd order Hall signal.

## Results

**Epitaxial FeSn films for electrical transport studies**

FeSn (space group $P6/mmm$, $a = 5.2959$ Å, and $c = 4.4481$ Å) belongs to the family of transition metal-based kagome metals, which has recently attracted much interest, owing to the presence of the kagome lattice derived flat bands [18–20], topological electronic states [6,19], and itinerant magnetism [21,22]. FeSn consists of individual $Fe_3Sn$ and stanene layers that are vertically stacked in alternating order along the crystallographic c-axis (Fig.1(a)). The iron (Fe) atoms are arranged on a 2D kagome lattice and contribute an approximate $2\mu_B$ magnetic moment per atom, which originates from the partially filled d-orbitals. Strong intra-layer magnetic exchange interactions $J$ between the spin $S = 1$ moments lead to in-plane ferromagnetism, whereas an inter-layer antiferromagnetic Heisenberg term $J_C \cong 0.1J$ results in layered A-type antiferromagnetic order along the c-axis with a Néel temperature $T_N = 365$ K [17,23,24]. Using solid phase epitaxy with a molecular beam epitaxy system [25], we have synthesized 33 nm thick FeSn thin films, which were shaped into Hall bar devices (Fig.1(b)) [26]. The high crystalline quality of the films, which nucleate in a single phase, was confirmed using X-ray diffraction, energy-dispersive X-ray spectroscopy, and transmission electron microscopy experiments (Fig.S1) [26]. Measurements of the temperature-dependent longitudinal resistivity $\rho_{xx}(T)$ demonstrate the film's metallic character with a residual resistance ratio $RRR = R(300\,K)/R(2\,K)$=23.9 (Fig.1(c)). The overall transport characteristics (Fig.S2) agree with previous studies of MBE-grown antiferromagnetic FeSn films [16,25]; the kink in $\rho_{xx}(T)$ at $T = 352$ K can be associated with the onset of AFM [27] at $T_N$, whose value for thin films was previously found to be slightly smaller compared to that of bulk crystals [25,27]. The comparably high Néel temperature combined with the ability to grow high-quality thin films makes FeSn particularly well suited to study nonlinear topological transport responses at room temperature.

**Measurement of anomalous Hall effects at room temperature**

We now turn to the measurement of the linear and nonlinear Hall effects. To this end, we apply a longitudinal a.c. current $I_x(\omega)$ to the Hall bar device D1 and measure the transverse Hall voltages $V_{xy}^{1\omega}$, $V_{xy}^{2\omega}$, and $V_{xy}^{3\omega}$ via phase-sensitive lock-in detection as a function of $B$ applied in the out-of-plane direction at room temperature $T = 300$ K (Fig.1(b)). Control measurements to ensure the accuracy of our lock-in detection method are discussed in the Methods section. In Fig.1(d), we display the anomalous contribution to the 1$^{st}$ order Hall resistivity $\rho_{AHE}^{1\omega}(B)$. $\rho_{AHE}^{1\omega}(B)$ can be obtained from the Hall resistivity $\rho_{xy}^{1\omega}(B)$ by subtracting the linear contribution to the Hall effect $\rho_L^{1\omega}$, $\rho_{AHE}^{1\omega}(B) = \rho_{xy}^{1\omega}(B) - \rho_L^{1\omega}(B)$ (*cf.* Fig.1(d) inset). $\rho_L^{1\omega}(B)$ probably contains contributions from the orbital Hall effect $R_H^{1\omega}$ and (an anomalous part linear in $B$) from the linear magnetic susceptibility of the Fe magnetic moments to the out-of-plane magnetic field $R_0^{1\omega}$ [17]. We note that their relative contributions to $\rho_L^{1\omega}$ are difficult to separate unless the out-of-plane magnetization is saturated. The resulting $\rho_{AHE}^{1\omega}(B)$ is an odd function of $B$ and saturates at large field values $|B| \gg 0$ T. By contrast, we do not observe a 2$^{nd}$ order AHE and $\rho_{AHE}^{2\omega}(B)$ fluctuates around zero resistivity (Fig.1(e)). On the other hand, we observe a strong 3rd order AHE (Fig.1(f)). We obtain the corresponding 3$^{rd}$ order anomalous Hall resistivity $\rho_{AHE}^{3\omega}(B)$ by subtracting the linear contribution $\rho_0^{3\omega}(B)$ from the 3$^{rd}$ order Hall resistivity $\rho_{xy}^{3\omega}(B)$, $\rho_{AHE}^{3\omega}(B) = \rho_{xy}^{3\omega}(B) - \rho_0^{3\omega}(B)$ (Fig.2(e) inset). As we will show below, the anomalous contribution to $\rho_{xy}^{3\omega}(B)$ predominantly arises from the Berry curvature quadrupole. While we focus our following analysis on the saturating part $\rho_{AHE}^{3\omega}(B)$ at small field $|B| < 1\,T$, $\rho_0^{3\omega}(B)$ possibly originates from the magnetic field induced orbital 3$^{rd}$ order Hall effect. Like $\rho_{AHE}^{\omega}(B)$, $\rho_{AHE}^{3\omega}(B)$ is an odd function of $B$ and displays $B$-independent characteristics at $|B| \gg 0$ T. We have further studied the dependence of the measured Hall voltages on the longitudinal current drive $I_x$ at $B = 8T$ (Fig.1, (g) and (h)). The observed $V_{xy}^{1\omega}(I_{xx}) \propto I_x$ and $V_{xy}^{3\omega}(I_x) \propto I_x^3$ relations establish the 1$^{st}$ and 3$^{rd}$ order nature of the measured Hall voltages.

**Experimental characterization of the Hall effects**

To characterize the observed AHEs in more detail, we studied their temperature dependence. Consistent with the room temperature observations, we do not observe a 2$^{nd}$ order AHE at any temperature (Fig.S3). In Fig.2(a), we display $\rho_{AHE}^{1\omega}(B)$ measured at different temperatures (Fig.S2, A and B show the temperature dependence of $\rho_{xy}^{1\omega}(B)$ and $\rho_{xy}^{3\omega}(B)$, respectively). While $\rho_{AHE}^{1\omega}(B)$ vanishes for $T \to 0$ K, the high field saturation value of $\rho_{AHE}^{1\omega}(B)$ monotonically increases with increasing temperature and plateaus at $T > 330$ K. By comparison, $\rho_{AHE}^{3\omega}(B)$ shows a much richer temperature response (Fig.2(b)). Like $\rho_{AHE}^{1\omega}(B)$, $\rho_{AHE}^{3\omega}(B)$ vanishes at $T \to 0$K and increases with temperature up to $T = 330$ K. However, $\rho_{AHE}^{3\omega}(B)$ then decreases and exhibits a sign change between 350 and 360 K. This contrasting behavior at $T > 330$ K is also illustrated in Fig.2, (c) and (d), which display $\rho_{AHE}^{1\omega}|_{B=-4\,T}(T)$ and $\rho_{AHE}^{3\omega}|_{B=-4\,T}(T)$. Furthermore, the 1$^{st}$ and 3$^{rd}$ order AHEs sensitively depend on the spatial orientation of $B$ with respect to the crystallographic axes of FeSn. When $B$ lies within the crystallographic $a$-$b$-plane, both $\rho_{AHE}^{1\omega}(B)$ and $\rho_{AHE}^{3\omega}(B)$ vanish in measurements conducted at $T = 300$ K (see Fig.2, (e) and (f); measurements at other temperatures are shown in Fig.S4). This concludes the main experimental results of our work, which demonstrates the 3$^{rd}$ order NLAHE over a wide temperature window extending beyond room temperature (see figs. S5 and S6 for measurements of other Hall bar devices D2 and D3, respectively). The observed 3$^{rd}$ order NLAHE in our study is an odd function of the magnetic field. Therefore, it must have a different origin than the Berry curvature dipole and Berry connection polarizability induced 2$^{nd}$ and 3$^{rd}$ order nonlinear Hall effects [10,28], which are an even function of the magnetic field, i.e., they are finite at zero magnetic field when $\mathcal{T}$ is intact. In the following, we show that our observations are consistent with a $Q_{\alpha\beta\gamma}$-induced 3$^{rd}$ order NLAHE in an A-type kagome antiferromagnet whose spins are canted toward the $c$-axis.

**Symmetry analysis and electronic structure calculations**

First, we illustrate this concept by considering the magnetic structure of FeSn (Fig.1(a)), which belongs to the magnetic point group $mmm1$'. The presence of finite Berry curvature moments follows from a symmetry analysis. Because $mmm1$' preserves $\mathcal{P} \cdot \mathcal{T}$, native FeSn does not exhibit a $\mathbf{\Omega}$-induced AHE. On the other hand, canting of the spins toward the $c$-axis, such as induced by an externally applied magnetic field, lowers the symmetry to 2/$m$, which breaks $\mathcal{P} \cdot \mathcal{T}$ and permits a finite $\mathcal{M}$ and $Q_{\alpha\beta\gamma}$, while $\mathcal{D}_{\alpha\beta}$ vanishes. Hence, an A-type kagome AFM with spin canting along the $c$-axis is expected to exhibit a 1$^{st}$ and 3$^{rd}$ AHE, whereas the 2$^{nd}$ order AHE is zero. We qualitatively determined the $\mathbf{\Omega}$ contributions to the linear and NLAHE in terms of a nearest-neighbour tight-binding model with spin-orbit coupling [26,29]. Fig.3, (a) and (b) display $\mathcal{M}$ and $Q_{xxz}$, respectively as a function of the out-of-plane spin canting $M_z$. While $\mathcal{M} = Q_{xxz} = 0$ if $M_z = 0$ ($\mathcal{P} \cdot \mathcal{T}$ preserved), $\mathcal{M}$ and $Q_{xxz}$ are finite if $M_z \neq 0$ ($\mathcal{P} \cdot \mathcal{T}$ broken). The underlying $\mathbf{\Omega}$ primarily originates from the massive Dirac cones at the $K$ points in the electronic structure (see Fig.S7). Even in the absence of AFM at $T > T_N$, $M_z \neq 0$ breaks $\mathcal{T}$ and results in $\mathcal{M}, Q_{xxz} \neq 0$ (see Fig.3, (c) and (d)). Notably, $Q_{xxz}$ changes sign at the transition from a canted antiferromagnetic to a canted paramagnetic phase at $T_N$. While not dictated by symmetry, this sign change highlights the sensitivity of $Q_{xxz}$ to the underlying magnetic and electronic material properties [30]. To further show that the realistic electronic structure of spin-canted FeSn supports a finite $Q_{xxz}$, we performed *ab-initio* calculations [26]. The resulting band structure of the density functional theory (DFT)-derived Wannier Hamiltonian in the antiferromagnetic phase (see Fig.S9) agrees with previous results [16] and reveals multiple electron and hole pockets at the Fermi level (see Fig.4, (a) and (b) insets for the calculated Fermi surfaces). We include a canting term $M_z$ to simulate the spin canting along the *c*-axis. At $M_z \neq 0$, the double band degeneracy in both the antiferromagnetic ($T < T_N$) and paramagnetic ($T > T_N$) state is lifted and yields finite $\mathcal{M}$ and $Q_{xxz}$ in the antiferromagnetic (Fig.4, (a) and (b)) and paramagnetic phase, respectively. We further estimate $Q_{xxz}$ from the resulting Wannier states and obtain $Q_{xxz,calc} \sim 10^0$ Å. Overall, our results of the symmetry analysis, tight-binding model, and the electronic structure calculation are consistent with our experimental observations and support a $Q_{xxz}$-induced 3rd order NLAHE.

**Spin-canted state of FeSn**

The spin-canted state of FeSn with a finite *c*-axis magnetization ($\sim 0.01 \mu_B$/Fe atom) in the presence of an out-of-plane magnetic field was previously established by magnetometry measurements over a wide temperature window (2 K to 300 K) [17]. Here, we use classical Monte Carlo simulations based on realistic magnetic exchange terms of FeSn [17] to further demonstrate that this canted state results from the interplay between critical thermal fluctuations near $T_N$ and an externally applied magnetic field [26]. The resulting in-plane $\langle M_{xy}(0,0,\pi)^2 \rangle$ and out-of-plane $\langle M_z^2 \rangle$ spin correlation functions are shown in Fig.4, (c) and (d), respectively. At $T \ll T_N$, the Fe spins lie almost parallel to the *a-b*-plane owing to a finite easy-plane crystalline anisotropy and exhibit A-type AFM order (*cf.* Fig. 1A). As temperature increases, thermal fluctuations gradually weaken this magnetic order and $\langle M_{xy}(0,0,\pi)^2 \rangle \to 0$ near $T_N = 0.76 J_\parallel$, which is consistent with the reported $T_N$ [17]. This "thermal softening" of the in-plane spin alignment permits a finite out-of-plane spin canting $\langle M_z^2 \rangle > 0$ in the presence of an out-of-plane magnetic field. This effect is the strongest near $T_N$ at which critical fluctuations render the spins extremely susceptible to external perturbations, in this case, the out-of-plane magnetic field. Therefore, $\langle M_z^2 \rangle$ gradually increases with temperature, peaks at $T_N$, and remains finite even at $T > T_N$. Magnetic-field-dependent simulations further confirm the linear susceptibility and magnitude of the previously experimentally determined out-of-plane magnetization (Fig.S10) [17]. Overall, these characteristics are consistent with our experimental observations of the 1st and 3rd order AHE both of which require an out-of-plane spin-canted state to take finite values: $\rho_{AHE}^{1\omega}$ and $\rho_{AHE}^{3\omega}$ vanish at $B = 0$ T and $T \to 0$ K, exhibit a monotonic growth toward $T_N$ in a finite out-of-plane magnetic field, and remain finite at $T > T_N$ and $B \neq 0$ T. We note that the residual $\langle M_z^2 \rangle$ at $B = 2$ T and $T = 0$ K could be too small to induce a measurable AHE. Possible spin-charge interactions [17], and quantum-mechanical corrections could further amend the low-temperature magnetic structure via second order effects not considered in our model.

## Discussion

**Excluding competing mechanisms**

It is important to distinguish intrinsic $Q_{xxz}$ contributions to the 3$^{rd}$ order AHE signal from possible extrinsic contributions, which can result from impurity scattering [31–33] and Joule heating [34,35]. First, our experimental observations show that $\rho_{AHE}^{3\omega}(B)$ changes sign at $T_N$ (Fig.2(d)) but $d\rho/dT$ does not change sign (Fig.1(c), inset). Hence, a Joule heating-induced 3$^{rd}$ order orbital Hall effect cannot explain the observed phenomenology of a sign-changing 3$^{rd}$ order NLAHE [35].

Second, to separate intrinsic from scattering-induced extrinsic contributions to the anomalous Hall signal, we developed a scaling law analysis for the 3$^{rd}$ order NLAHE in terms of the charge carrier scattering time $\tau$ [13,14,26,36]. To this end, we analyze the temperature dependence of the anomalous Hall ($E_{AHE}^{3\omega}$) and longitudinal ($E_{xx}$) electric field ratio $E_{AHE}^{3\omega}/E_{xx}^3$ and the longitudinal conductivity $\sigma_{xx}$ at $T < T_N$ (Fig.5(a)) [26]. The scattering time dependence can be parametrized as $E_{AHE}^{3\omega}/(E_{xx}^3 \sigma_{xx}) = \alpha \sigma_{xx}^2 + \beta$, where $\alpha$ accounts for skew scattering and $\beta = \frac{m_{eff}^2}{2\hbar^3 n^2} Q_{xxz}$ for the intrinsic contribution [36] ($m_{eff}$ denotes the effective electron mass, $\hbar$ the reduced Planck's constant, and $n$ the charge carrier density). Because the intrinsic contribution is independent of $\tau$, it dominates the Hall response in the limit $\sigma_{xx}^2 \to 0$ and is proportional to the vertical intercept $\beta$. Indeed, our corresponding analysis (Fig.5(b)) demonstrates that intrinsic contributions dominate the 3$^{rd}$ order AHE over a wide temperature range (approximately 95% at $T = 300$ K), while skew scattering only becomes relevant at $T < 100$ K. This distinguishes our results from existing work on nonlinear Hall effects in which the relative contribution of the topological transport response to the AHE is much smaller owing to the prevalence of skew scattering [14,37]. Importantly, our analysis demonstrates the universal $\tau^4$ −scaling of the skew scattering contribution to the 3$^{rd}$ order Hall effect. Using $\beta = (7.6 \pm 0.2) \times 10^2$ $\Omega\mu m^3 V^{-2}$ obtained from fitting $E_{AHE}^{3\omega}/(E_{xx}^3 \sigma_{xx})$, $m_{eff} = 5.4$ $m_e$ [16] ($m_e$ denotes the electron mass), and $n = 1 \times 10^{21}$ cm$^{-3}$ (from density functional theory calculations [26]), we obtain $Q_{xxz,exp} = 73 \pm 2$ Å. We note that the equivalent scaling analysis [36] of the 1$^{st}$ order AHE (Fig.S8) also suggests an intrinsic origin of the observed Hall signal, consistent with our expectation from the symmetry analysis and electronic structure calculations. The deviation of $Q_{xxz,exp}$ from the calculated estimate $Q_{xxz,calc} \sim 10^0$ Å possibly originates from a limited accuracy of DFT in obtaining the realistic electronic structure of FeSn and a complex Fermi surface (Fig.4, (a) and (b) insets) [16], which renders the momentum space integration of the narrow $\mathbf{\Omega}$- and $\partial_x \partial_x \mathbf{\Omega}$-distributions challenging [26]. Moreover, uncertainties in the actual $m_{eff}$ [16], as well as an impurity induced reduction of $n$ could further influence the $Q_{xxz,exp}$ magnitude [26].

Finally, while the multiband Fermi surface of FeSn (Fig.4, (a) and (b) insets) could generally contribute to an S-shaped Hall response (Fig.1(f)), the derived scaling law is robust against multiband transport [26]. Hence, based on the presented scaling analysis, we can exclude multiband transport as a possible origin of the observed 3$^{rd}$ order NLAHE. Because we are not aware of other mechanisms that can give rise to a 3$^{rd}$ order Hall signal with these scaling characteristics, our analysis strongly supports the Berry curvature quadrupole origin of the observed 3$^{rd}$ order NLAHE.

**Conclusion**

Our study presents experimental evidence for the Berry curvature quadrupole induced $3^{rd}$ order NLAHE at room temperature in spin-canted FeSn [12,30]. This conclusion is supported by a scaling law analysis, a symmetry analysis, model calculations, first-principle calculations, and magnetic Monte-Carlo simulations whose results are consistent with the experimentally observed phenomenology. Hence, our work suggests that Berry curvature multipoles can be supported by the magnetic point group symmetries of antiferromagnets, which is a fundamentally new mechanism. Because Berry curvature multipoles are predicted to exist for ninety magnetic point groups [12], our work opens a new field to study a variety of magnetic materials, whose topological electronic states were previously inaccessible, through nonlinear measurement protocols. Owing to the observed sensitivity of the $3^{rd}$ order NLAHE to the electronic and magnetic material properties [12,30], our work showcases the use of NLAHEs as a sensitive electric transport probe to investigate antiferromagnetic phase transitions of various materials, such as moiré superlattices [38], two-dimensional van der Waal magnets [39], and quantum spin liquid materials [40,41] that remain poorly understood to date and that are notoriously difficult to study by using electric transport measurements. Our work further extends previous research on AHEs in antiferromagnets [42–44] to the nonlinear transport regime and highlights antiferromagnets [45–47] as promising candidates to exhibit useful quantum properties at room temperature, owing to comparably large antiferromagnetic exchange terms, the absence of a large net magnetization, and the rich magnetic symmetries. In this regard, the observation of the $3^{rd}$ order NLAHE in an epitaxially grown kagome magnet [25,48,49] in our study narrows the gap toward the technological use of nonlinear Hall effects in high-frequency and energy-harvesting applications [50].

## Methods
**Sample Preparation**
Epitaxial FeSn thin films were grown on SrTiO₃ (STO) substrates in (111)-direction (from CrysTec GmbH) by using a home-built molecular beam epitaxy (MBE) setup. The as-received STO (111) substrates were cleaned using sonication in acetone and iso propyl alcohol (IPA) for 5 min each. Single-terminated STO (111) surfaces were obtained by using a hot water etching step performed at 90° C for 90 minutes. Thermal annealing of the STO (111) substrates was performed at 1050° C with an oxygen flow of 50 sccm inside a tube furnace for 1 hour. Prior to the thin film growth, substrates were outgassed inside the MBE chamber at 600° C at a base pressure of $< 5 \times 10^{-10}$ mbar. Following the growth recipe of Inoue *et al.* [25], (high purity Fe (99.99%) and Sn (99.99%) (Alfra Aesar) were co-evaporated from Knudsen effusion cells at a beam flux ratio of 1:2.2 at an approximate growth rate of 1 nm/min to obtain a 30 nm thick amorphous FeSn film. During the deposition, the STO (111) substrate was held at $T = 200°$ C. Onwards, the films were capped with an approximately 6 nm thick layer of amorphous BaF₂. FeSn of high crystalline quality were obtained by *in-situ* post-annealing of the deposited FeSn layer at T= 500° C using a ramp rate of 5° C/min during the heat up and cool down phase. The FeSn thin films were patterned into six terminal Hall bars (Fig. 1(b)) by using Ar⁺ ion milling and optical UV lithography. During the whole fabrication process, the FeSn films are protected by the BaF₂ capping layer. Electrical contacts to the FeSn Hall bar structure are fabricated by evaporating 5 nm/100 nm titanium/gold electrodes. We note that cleaved [51] and argon ion irradiated [52] STO (111) surfaces can exhibit conducting two-dimensional electron gases at the substrate surface. Neither prepared STO substrates prior FeSn deposition nor STO substrates after argon ion milling of the Hall bar structures exhibited a finite conductance across the substrate surface in four-probe measurements conducted at ambient conditions. This indicates the absence of substrate surface effects in our electric transport measurements.

**Electric transport measurements**
All transport measurements were carried out in a commercial cryogenic magnet system (J4804 from Cryogenic Limited) using a four-probe contact geometry. The Hall bar devices were wire bonded with 40 μm aluminum wire to the chip carrier. Two-terminal measurements between different contact pairs were carried out to ensure Ohmic device contacts (Fig.S11). An a.c. current drive of $I_x(\omega) = I_0 \sin(\omega t)$ with a peak amplitude peak of $I_0 = 5$ mA at a frequency of $f = \frac{\omega}{2\pi} = 19.357$ Hz was applied by using a source meter (6221A from Keithley). For DC measurements, an $I_x = 100$ μA d.c current bias was applied. Four SR830 (Stanford Research Systems) lock-in amplifiers, which are phase matched to the $I_x$ output, were used to simultaneously measure $V_{xx}^{1\omega}$, $V_{xy}^{1\omega}$, $V_{xy}^{2\omega}$, and $V_{xy}^{3\omega}$ (Fig.1(b)). Both $V_{xy}^{\omega}$ and $V_{xy}^{3\omega}$ are in phase with the drive frequency, whereas $V_{xy}^{2\omega}$ is out of phase by π/2 with respect to the drive frequency. The lock-in phase was calibrated and monitored for each measurement. Typical phase data of $V_{xy}^{\omega}$ and $V_{xy}^{3\omega}$ are shown in Fig.S12. Measurements of $V_{xy}^{1\omega}$ were conducted using different filters, which have different roll-off characteristics, to ensure the spectral purity of the measured lock-in signals (see Fig.S13). Before any data analysis presented in the main text was conducted, zero field constant voltage offsets were removed from $V_{xy}^{\omega}$, $V_{xy}^{2\omega}$, and $V_{xy}^{3\omega}$. These offsets likely arise from a slight misalignment between the FeSn Hall bar and the electrodes causing a finite albeit small coupling of the longitudinal signal into the Hall voltages. The zero-field offset Hall signal was found to exhibit random values across different devices. The longitudinal $\rho_{xx}^{1\omega} = \left(\frac{dL_{xy}}{L_{xx}}\right)\frac{V_{xx}^{1\omega}}{I_x}$ and 1ˢᵗ and 3ʳᵈ order Hall $\rho_{xy}^{1\omega,3\omega} =$ $\left(\frac{dL_{xx}}{L_{xy}}\right)\frac{V_{xx}^{1\omega,3\omega}}{I_x}$ resistivities were obtained by considering the film thickness $d = 30$ nm and Hall bar dimensions $L_{xx} = 26$ μm and $L_{xy} = 21$ μm.

# References

[1] R. Karplus and J. M. Luttinger, *Hall Effect in Ferromagnetics*, Phys. Rev. **95**, 1154 (1954).
[2] F. D. M. Haldane, *Berry Curvature on the Fermi Surface: Anomalous Hall Effect as a Topological Fermi-Liquid Property*, Phys Rev Lett **93**, 206602 (2004).
[3] N. Nagaosa, J. Sinova, S. Onoda, A. H. MacDonald, and N. P. Ong, *Anomalous Hall Effect*, Rev Mod Phys **82**, 1539 (2010).
[4] E. Liu et al., *Giant Anomalous Hall Effect in a Ferromagnetic Kagome-Lattice Semimetal*, Nat Phys **14**, 1125 (2018).
[5] Q. Wang, Y. Xu, R. Lou, Z. Liu, M. Li, Y. Huang, D. Shen, H. Weng, S. Wang, and H. Lei, *Large Intrinsic Anomalous Hall Effect in Half-Metallic Ferromagnet Co3Sn2S2 with Magnetic Weyl Fermions*, Nature Communications 2018 9:1 **9**, 1 (2018).
[6] L. Ye et al., *Massive Dirac Fermions in a Ferromagnetic Kagome Metal*, Nature 2018 555:7698 **555**, 638 (2018).
[7] C.-Z. Chang et al., *Experimental Observation of the Quantum Anomalous Hall Effect in a Magnetic Topological Insulator*, Science (1979) **340**, 167 (2013).
[8] M. Serlin, C. L. Tschirhart, H. Polshyn, Y. Zhang, J. Zhu, K. Watanabe, T. Taniguchi, L. Balents, and A. F. Young, *Intrinsic Quantized Anomalous Hall Effect in a Moiré Heterostructure*, Science (1979) **367**, 900 (2020).
[9] Y. Deng, Y. Yu, M. Z. Shi, Z. Guo, Z. Xu, J. Wang, X. H. Chen, and Y. Zhang, *Quantum Anomalous Hall Effect in Intrinsic Magnetic Topological Insulator MnBi2Te4*, Science (1979) **367**, 895 (2020).
[10] I. Sodemann and L. Fu, *Quantum Nonlinear Hall Effect Induced by Berry Curvature Dipole in Time-Reversal Invariant Materials*, Phys Rev Lett **115**, 216806 (2015).
[11] H. Liu, J. Zhao, Y. X. Huang, W. Wu, X. L. Sheng, C. Xiao, and S. A. Yang, *Intrinsic Second-Order Anomalous Hall Effect and Its Application in Compensated Antiferromagnets*, Phys Rev Lett **127**, 277202 (2021).
[12] C. P. Zhang, X. J. Gao, Y. M. Xie, H. C. Po, and K. T. Law, *Higher-Order Nonlinear Anomalous Hall Effects Induced by Berry Curvature Multipoles*, Phys Rev B **107**, 115142 (2023).
[13] Q. Ma et al., *Observation of the Nonlinear Hall Effect under Time-Reversal-Symmetric Conditions*, Nature 2018 565:7739 **565**, 337 (2018).
[14] K. Kang, T. Li, E. Sohn, J. Shan, and K. F. Mak, *Nonlinear Anomalous Hall Effect in Few-Layer WTe2*, Nature Materials 2019 18:4 **18**, 324 (2019).
[15] M. Huang et al., *Intrinsic Nonlinear Hall Effect and Gate-Switchable Berry Curvature Sliding in Twisted Bilayer Graphene*, Phys Rev Lett **131**, 066301 (2023).
[16] M. Kang et al., *Dirac Fermions and Flat Bands in the Ideal Kagome Metal FeSn*, Nat Mater **19**, 163 (2020).
[17] B. C. Sales, J. Yan, W. R. Meier, A. D. Christianson, S. Okamoto, and M. A. McGuire, *Electronic, Magnetic, and Thermodynamic Properties of the Kagome Layer Compound FeSn*, Phys Rev Mater **3**, 114203 (2019).
[18] W. R. Meier, M.-H. Du, S. Okamoto, N. Mohanta, A. F. May, M. A. McGuire, C. A. Bridges, G. D. Samolyuk, and B. C. Sales, *Flat Bands in the CoSn-Type Compounds*, Phys Rev B **102**, 075148 (2020).
[19] M. Kang et al., *Topological Flat Bands in Frustrated Kagome Lattice CoSn*, Nat Commun **11**, 4004 (2020).
[20] C. Chen, J. Zheng, R. Yu, S. Sankar, H. C. Po, K. T. Law, and B. Jäck, *Visualizing the Localized Electrons of a Kagome Flat Band*, (2023).
[21] B. C. Sales, W. R. Meier, D. S. Parker, L. Yin, J. Q. Yan, A. F. May, S. Calder, A. A. Aczel, Q. Zhang, and H. Li, *Flat-Band Itinerant Antiferromagnetism in the Kagome Metal CoSn1-XInx*, ArXiv Preprint ArXiv:2201.12421 (2022).
[22] L. Ye et al., *A Flat Band-Induced Correlated Kagome Metal*, ArXiv Preprint ArXiv:2106.10824 (2021).
[23] L. O. Vogt et al., *Studies of the Magnetic Structure of FeSn Using the Mössbauer Effect*, Phys Scr **11**, 47 (1975).
[24] Y. Xie, L. Chen, T. Chen, Q. Wang, Q. Yin, J. R. Stewart, M. B. Stone, L. L. Daemen, E. Feng, and H. Cao, *Spin Excitations in Metallic Kagome Lattice FeSn and CoSn*, Commun Phys **4**, 1 (2021).
[25] H. Inoue, M. Han, L. Ye, T. Suzuki, and J. G. Checkelsky, *Molecular Beam Epitaxy Growth of Antiferromagnetic Kagome Metal FeSn*, Appl Phys Lett **115**, 72403 (2019).
[26] *See Supplementary Materials.*
[27] L. O. Vogt, C. Villevieille -Chemical, E. Properties of FeSn, R. A. Covert, H. H. Uhlig -, H. Lu, Y. Liu, X. Kou -, and B. Stenstrom, *The Electrical Resistivity of FeSn Single Crystals*, Phys Scr **6**, 214 (1972).

[28] Y. Gao, S. A. Yang, and Q. Niu, *Field Induced Positional Shift of Bloch Electrons and Its Dynamical Implications*, Phys Rev Lett **112**, 166601 (2014).

[29] H. M. Guo and M. Franz, *Topological Insulator on the Kagome Lattice*, Phys Rev B Condens Matter Mater Phys **80**, 113102 (2009).

[30] X. Zhang, K. Sun, and Z. Y. Meng, *The "Sign Problem" of the 3rd Order Anomalous Hall Effect in Topological Magnetic Materials*, (2023).

[31] J. Smit, *The Spontaneous Hall Effect in Ferromagnetics I*, Physica **21**, 877 (1955).

[32] J. Smit, *The Spontaneous Hall Effect in Ferromagnetics II*, Physica **24**, 39 (1958).

[33] L. Berger, *Side-Jump Mechanism for the Hall Effect of Ferromagnets*, Phys. Rev. B **2**, 4559 (1970).

[34] C. Dames and G. Chen, *1ω,2ω, and 3ω Methods for Measurements of Thermal Properties*, Review of Scientific Instruments **76**, 124902 (2005).

[35] X. Huang, C. Guo, C. Putzke, J. Diaz, K. Manna, C. Shekhar, C. Felser, and P. J. W. Moll, *Non-Linear Shubnikov-de Haas Oscillations in the Self-Heating Regime*, Appl Phys Lett **119**, 224101 (2021).

[36] Y. Tian, L. Ye, and X. Jin, *Proper Scaling of the Anomalous Hall Effect*, Phys Rev Lett **103**, (2009).

[37] S. Lai et al., *Third-Order Nonlinear Hall Effect Induced by the Berry-Connection Polarizability Tensor*, Nature Nanotechnology 2021 16:8 **16**, 869 (2021).

[38] Y. Tang et al., *Simulation of Hubbard Model Physics in WSe2/WS2 Moiré Superlattices*, Nature 2020 579:7799 **579**, 353 (2020).

[39] Y. Xu et al., *Coexisting Ferromagnetic–Antiferromagnetic State in Twisted Bilayer CrI3*, Nature Nanotechnology 2021 17:2 **17**, 143 (2021).

[40] J. A. Sears, M. Songvilay, K. W. Plumb, J. P. Clancy, Y. Qiu, Y. Zhao, D. Parshall, and Y. J. Kim, *Magnetic Order in α - RuCl3: A Honeycomb-Lattice Quantum Magnet with Strong Spin-Orbit Coupling*, Phys Rev B Condens Matter Mater Phys **91**, 144420 (2015).

[41] Y. Kasahara et al., *Majorana Quantization and Half-Integer Thermal Quantum Hall Effect in a Kitaev Spin Liquid*, Nature **559**, 227 (2018).

[42] C. Sürgers, G. Fischer, P. Winkel, and H. V. Löhneysen, *Large Topological Hall Effect in the Non-Collinear Phase of an Antiferromagnet*, Nature Communications 2014 5:1 **5**, 1 (2014).

[43] X. Cao et al., *Giant Nonlinear Anomalous Hall Effect Induced by Spin-Dependent Band Structure Evolution*, Phys Rev Res **4**, 023100 (2022).

[44] Z. Feng et al., *An Anomalous Hall Effect in Altermagnetic Ruthenium Dioxide*, Nature Electronics 2022 5:11 **5**, 735 (2022).

[45] A. K. Nayak et al., *Large Anomalous Hall Effect Driven by a Nonvanishing Berry Curvature in the Noncolinear Antiferromagnet Mn3Ge*, Sci Adv **2**, (2016).

[46] L. Šmejkal, A. H. MacDonald, J. Sinova, S. Nakatsuji, and T. Jungwirth, *Anomalous Hall Antiferromagnets*, Nature Reviews Materials 2022 7:6 **7**, 482 (2022).

[47] S. Nakatsuji, N. Kiyohara, and T. Higo, *Large Anomalous Hall Effect in a Non-Collinear Antiferromagnet at Room Temperature*, Nature **527**, 212 (2015).

[48] D. Hong, C. Liu, H.-W. Hsiao, D. Jin, J. E. Pearson, J.-M. Zuo, and A. Bhattacharya, *Molecular Beam Epitaxy of the Magnetic Kagome Metal FeSn on LaAlO3 (111)*, AIP Adv **10**, 105017 (2020).

[49] S. Cheng, B. Wang, I. Lyalin, N. Bagués, A. J. Bishop, D. W. McComb, and R. K. Kawakami, *Atomic Layer Epitaxy of Kagome Magnet Fe3Sn2 and Sn-Modulated Heterostructures*, APL Mater **10**, 061112 (2022).

[50] D. Kumar, C. H. Hsu, R. Sharma, T. R. Chang, P. Yu, J. Wang, G. Eda, G. Liang, and H. Yang, *Room-Temperature Nonlinear Hall Effect and Wireless Radiofrequency Rectification in Weyl Semimetal TaIrTe4*, Nature Nanotechnology 2021 16:4 **16**, 421 (2021).

[51] L. Miao, R. Du, Y. Yin, and Q. Li, *Anisotropic Magneto-Transport Properties of Electron Gases at SrTiO3 (111) and (110) Surfaces*, Appl Phys Lett **109**, 261604 (2016).

[52] S. McKeown Walker, A. De La Torre, F. Y. Bruno, A. Tamai, T. K. Kim, M. Hoesch, M. Shi, M. S. Bahramy, P. D. C. King, and F. Baumberger, *Control of a Two-Dimensional Electron Gas on SrTiO3 (111) by Atomic Oxygen*, Phys Rev Lett **113**, 177601 (2014).

**Acknowledgements:**
We acknowledge fruitful discussions with Xi Dai, Junwei Liu, and Hisashi Inoue. We appreciate the support of Hui Li, Chenjie Zhou, Jiannong Wang, and Wilwin at an early stage of this project. This work was primarily supported by the Hong Kong RGC (No. 26304221, B.J.) and the Croucher foundation (B.J.), the National Key R&D Program of China (Grant No. 2021YFA1401500, R.L., K.Q., and Q.S), and the Hong Kong RGC (Grant Nos. 17301420, 17301721, AoE/P-701/20, 17309822, A_HKU703/22, Z.Y.M.). Furthermore, B.J. acknowledges financial support by the start-up fund of the Hong Kong University of Science and Technology, Z.Y.M. acknowledges the K. C. Wong Education Foundation (Grant No. GJTD-2020-01), K.T.L. acknowledges the support of The Ministry of Science and Technology, China and HKRGC through 2020YFA0309601, RFS2021-6S03, C6025-19G, AoE/P-701/20, 16310520 and 16310219. Q.-F.L. acknowledges financial support of the MERIT-WINGS course provided by the University of Tokyo, and the Fellowship for Integrated Materials Science and Career Development provided by the Japan Science and Technology Agency. C.C. acknowledges support through a postdoctoral fellowship by the Tin Ka Ping foundation. Q.-F.L. further acknowledge support by the computational resource of Fujitsu PRIMERGY CX2550M5/CX2560M5 (Oakbridge-CX) awarded by "Large-scale HPC Challenge" Project, Information Technology Center, The University of Tokyo, for the Monte Carlo simulation.

**Author contributions:** S.S., J.Z., Y.H.L. grew the thin film samples under the supervision of C.C. R.L. carried out the electrical transport measurements. C.Z., X.G., and R.Y. carried out the tight-binding model and DFT calculations. Q.F.L and X.Z. carried out the Monte Carlo simulations. K.Q. fabricated the Hall bar devices. S.S., R.L., and C.Z. analysed the data. B.J. conceived the study. B.J., Q.S., K.T.L., and Z.Y.M. supervised the study. All authors discussed the results and contributed to the manuscript, which was written by B.J.

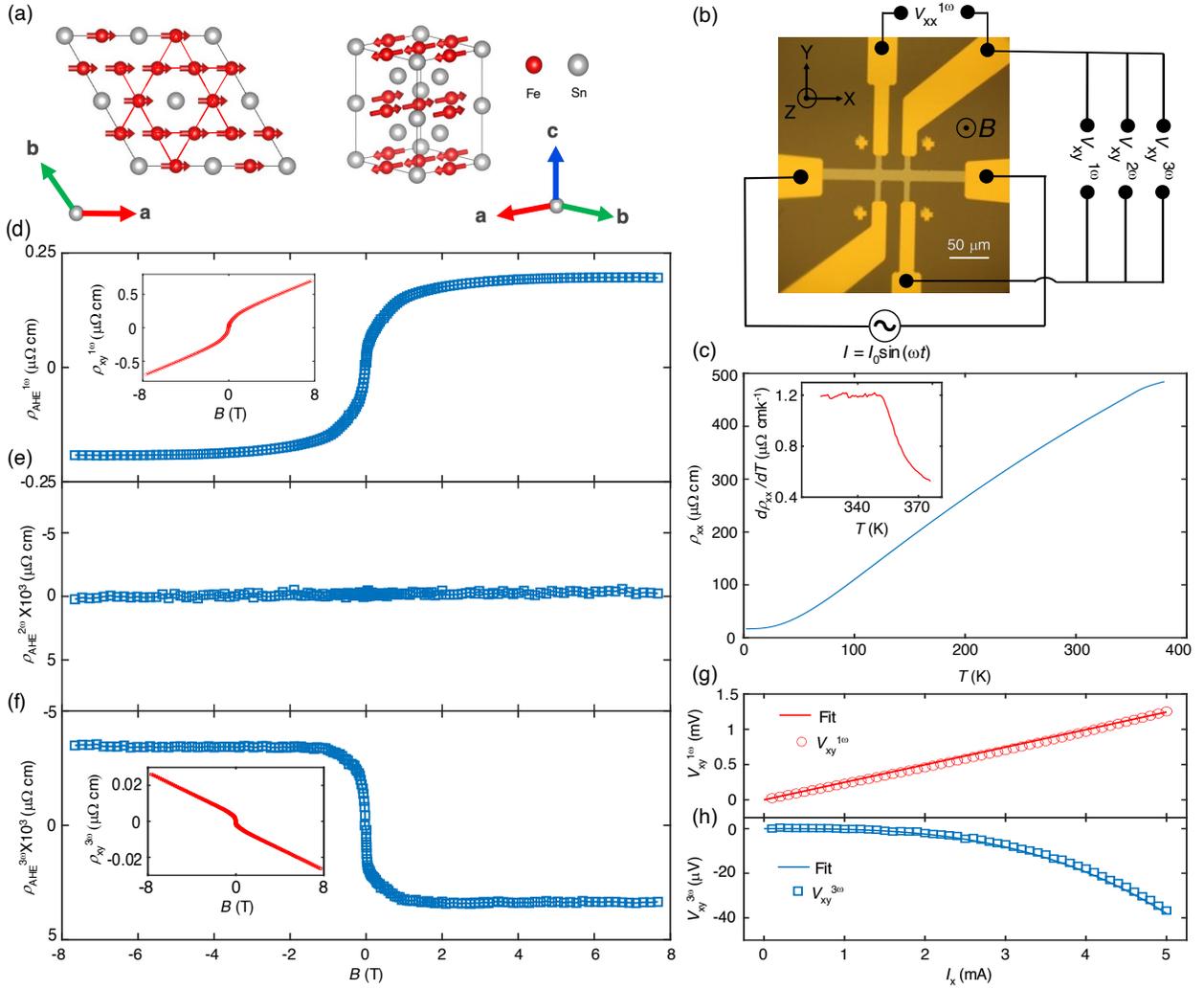

**Figure 1: Room temperature 3rd order nonlinear anomalous Hall effect in FeSn thin films.** (a) Crystallographic and magnetic structure of FeSn forming an A-type antiferromagnet. The iron (Fe) atoms occupy the sites of the kagome lattice that coordinates a triangular tin (Sn) lattice. The vertical stacking of the kagome planes along the $c$-axis is coordinated by individual stanene spacer layers. (b) Optical micrograph of the Hall bar device and the electrical transport measurement geometry [26]. (c) Measurement of the temperature $T$ dependent longitudinal resistivity $\rho_{xx}(T)$. The inset displays the corresponding derivative $d\rho_{xx}(T)/dT$. (d) Linear $\rho_{AHE}^{1\omega}$, (e) 2nd order nonlinear $\rho_{AHE}^{2\omega}$, and (f) 3rd order nonlinear $\rho_{AHE}^{3\omega}$ anomalous Hall resistivity as a function of the magnetic field $B$ measured at $T = 300$ K. The insets in panels a and f display the corresponding Hall resistivities $\rho_{xy}^{1\omega}(B)$ and $\rho_{xy}^{3\omega}(B)$, respectively before the linear background subtraction (see main text). (g) 1st order linear $V_{xy}^{1\omega}$, and (h) 3rd order nonlinear $V_{xy}^{3\omega}$ Hall voltages measured as a function of the longitudinal drive current $I_x$ at $T = 300$ K and $B = 8$ T (open symbols). The solid lines show a linear and cubic fit, respectively to the experimental data.

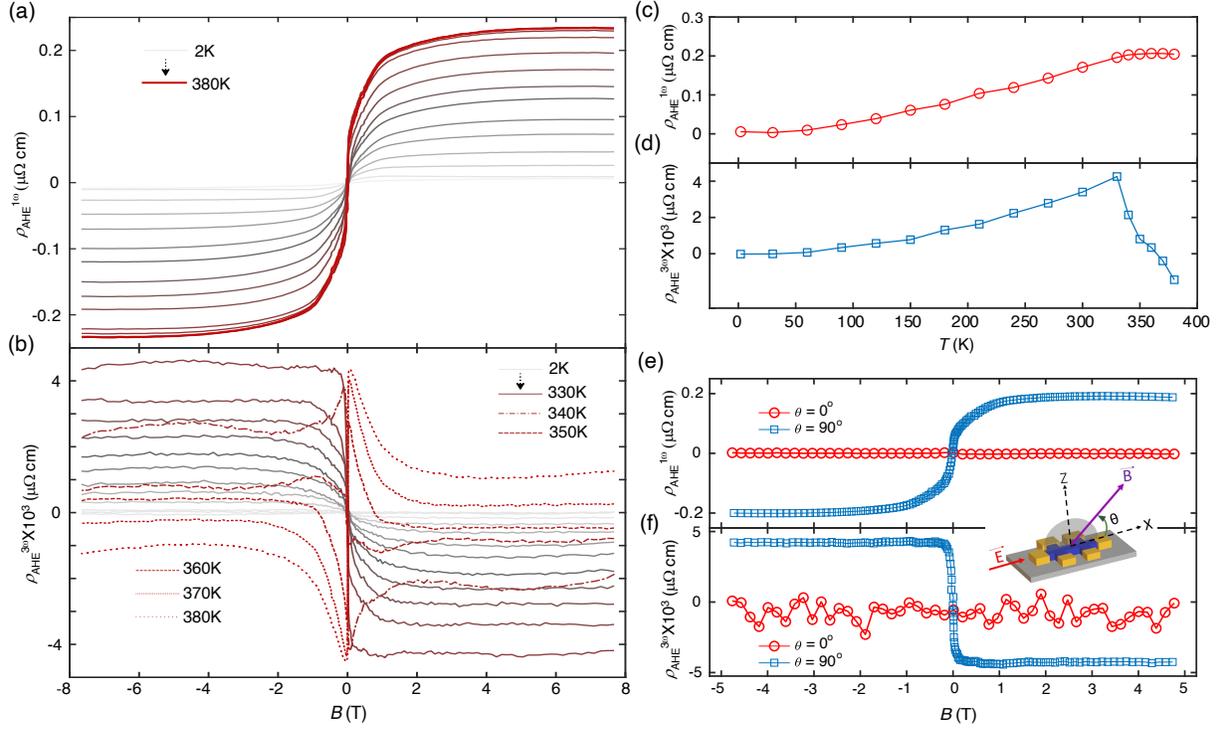

**Figure 2: Experimental characterization of the linear and nonlinear anomalous Hall effects.**
(a) Linear $\rho_{AHE}^{1\omega}$ and (b) 3$^{rd}$ order nonlinear $\rho_{AHE}^{3\omega}$ anomalous Hall resistivities as a function of the magnetic field $B$ measured at different temperatures $T$. (c) $\rho_{AHE}^{1\omega}$ and (d) $\rho_{AHE}^{3\omega}$ at $B = -4$ T plotted as a function of $T$. (e) $\rho_{AHE}^{1\omega}(B)$ and (f) $\rho_{AHE}^{3\omega}(B)$ measured at different spatial orientations of $B$ with respect to the longitudinal Hall bar axis $X$. At $\Theta = 0°$, $B$ is aligned in parallel to $X$ and at $\Theta = 90°$, $B$ is aligned perpendicular to $X$ in the out-of-plane direction that is parallel to the crystallographic $c$-axis. In all measurements, the current bias $I_x$ is applied along the $X$ direction.

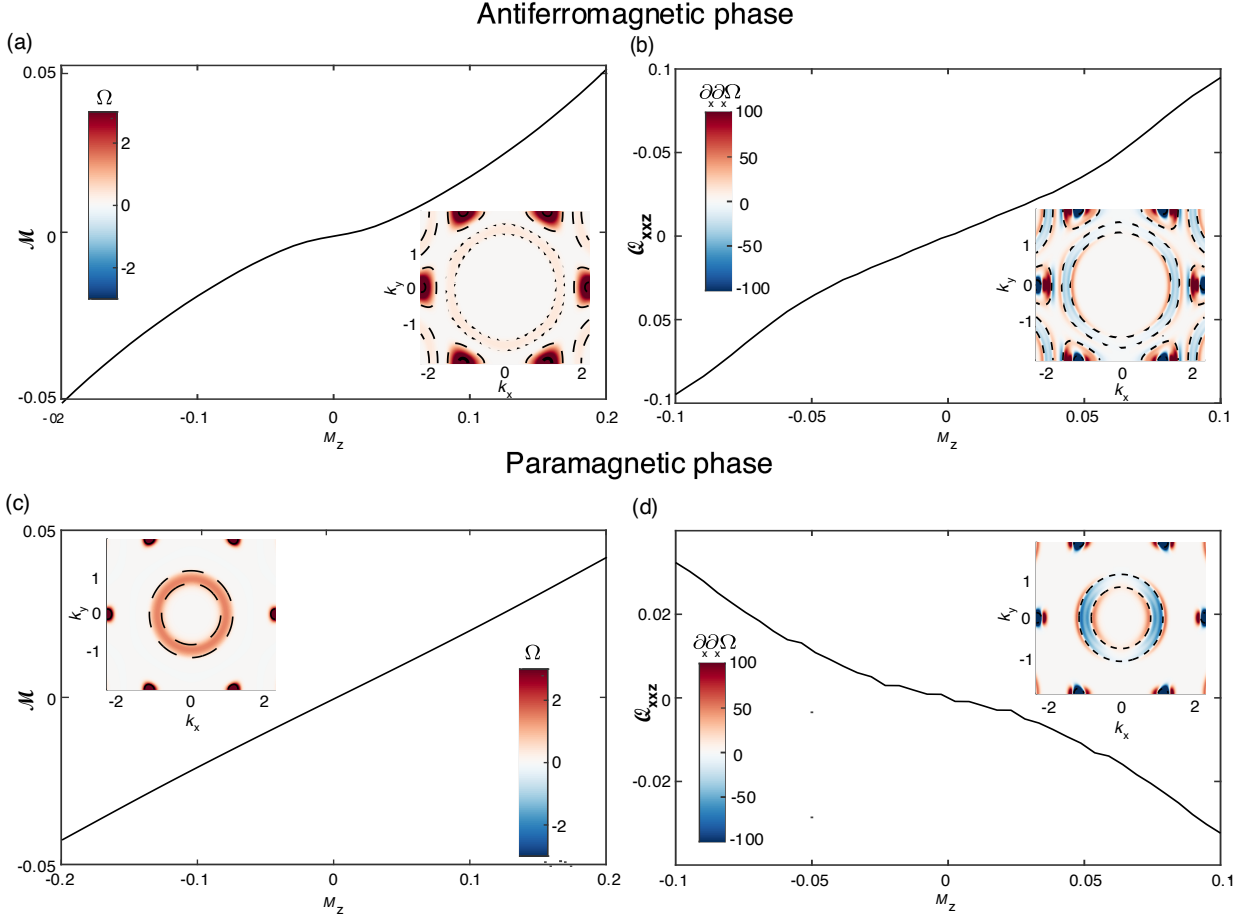

**Figure 3: Berry curvature quadrupole of an A-type antiferromagnet on the kagome lattice with canted spins in the antiferromagnetic and paramagnetic phase.** (a) Berry curvature monopole $\mathcal{M}$ and (b) Berry curvature quadrupole $\mathcal{Q}_{xxz}$ of an A-type kagome antiferromagnet as a function of an out-of-plane magnetization $M_z$, owing to spin canting, as calculated from the a tight-binding lattice model. The corresponding momentum $k_{x,y}$-dependent Berry curvature $\Omega$ and spatial derivative $\partial_x\partial_x\Omega$ are shown in the insets. (c) $\mathcal{M}$ and (d) $\mathcal{Q}_{xxz}$ of the canted kagome paramagnet as a function of $M_z$. The corresponding momentum $k_{x,y}$-dependent $\Omega$ and $\partial_x\partial_x\Omega$ are shown in the respective insets. See main text and Ref. [26] for model details.

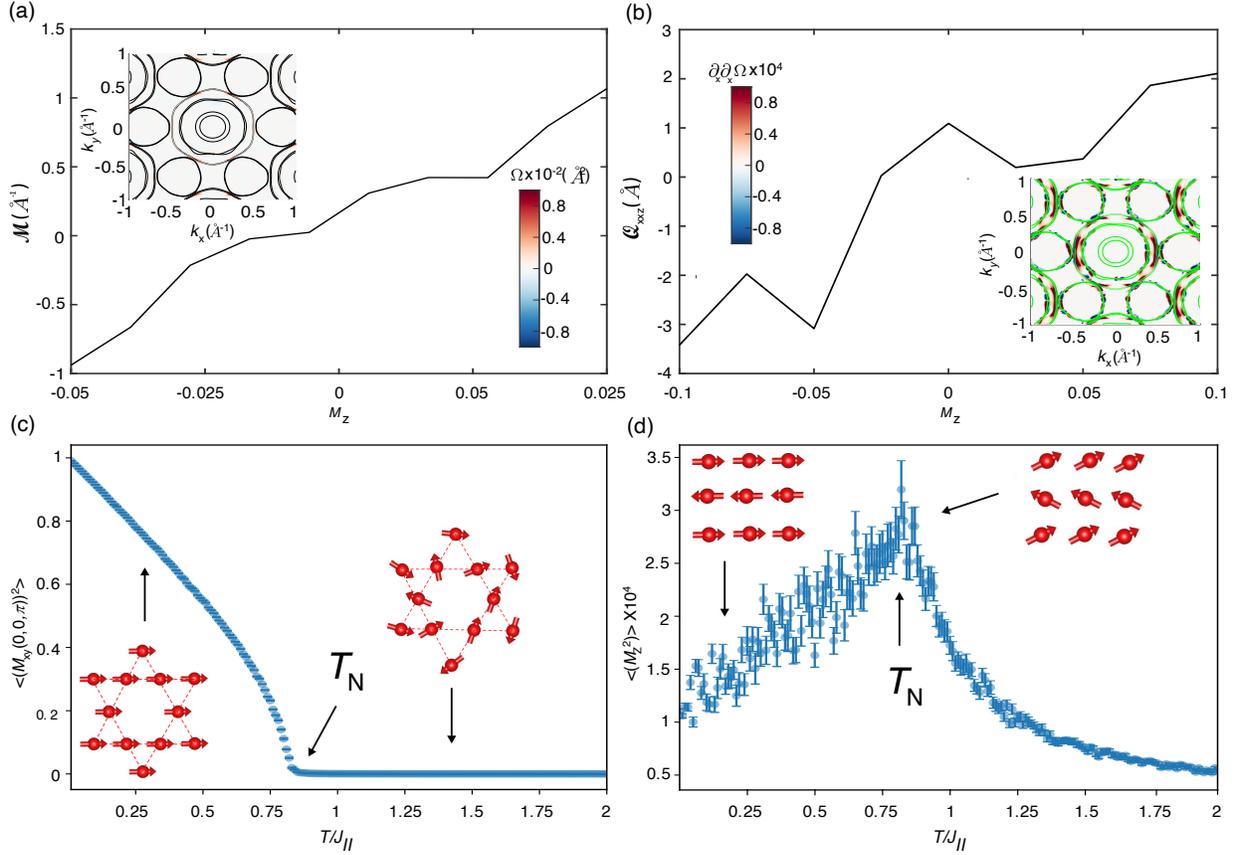

**Figure 4: Berry curvature monopole and quadrupole in the spin-canted state of FeSn**. (a) Berry curvature monopole $\mathcal{M}$ as a function of the out-of-plane spin canting $M_z$, calculated from the DFT-derived electronic structure of FeSn [16,26]. The inset displays the corresponding momentum $k_{x,y}$-dependent Berry curvature $\Omega$ distribution at Fermi energy ($M_z = 0.05$). (b) Corresponding Berry curvature quadrupole $\mathcal{Q}_{xxz}$ as a function of $M_z$. The inset displays the corresponding $k_{x,y}$-dependent second derivative $\partial_x \partial_x \Omega$ at Fermi energy and the Fermi surface is depicted as a green line. (c) In-plane $\langle M_{xy}(0,0,\pi)^2 \rangle$ and (c) out-of-plane $\langle M_z^2 \rangle$ spin correlations as a function of temperature $T$ in the presence of an external magnetic field $B = 2$ T applied along the crystallographic c-axis as obtained from magnetic Monte Carlo simulations [26]. $T$ is parametrized in terms of the ferromagnetic in-plane exchange $J_\parallel$. The insets in panel (a) schematically display the corresponding in-plane spin-polarization both in the ordered state near $T = 0$ K and above Néel temperature $T_N$ when in-plane spin-correlations vanish. The insets in panel (b) schematically display the corresponding out-of-plane spin-polarization at $T \ll T_N$ when spin canting is weak and at $T \leq T_N$ when the canting effect is maximum.

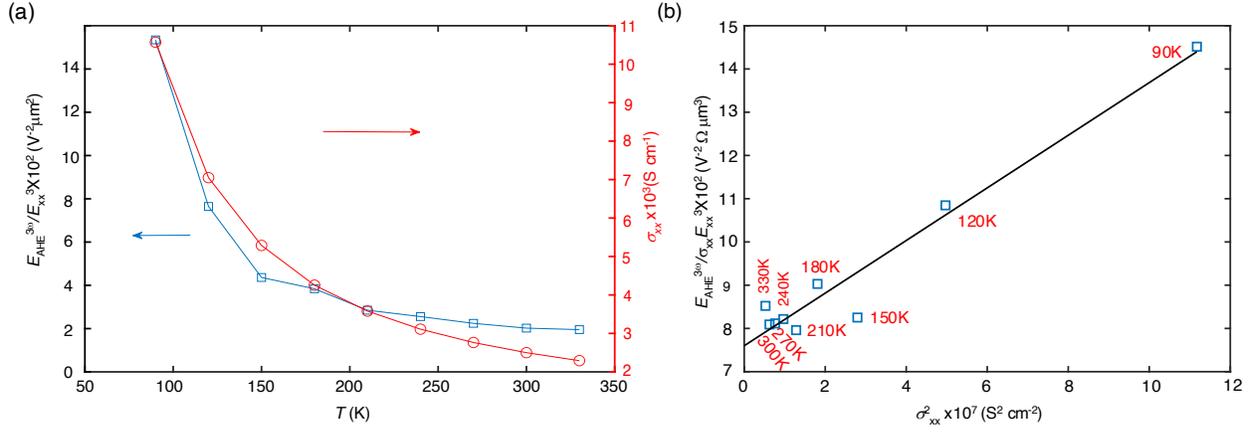

**Figure 5: Scaling law analysis of the 3rd order nonlinear AHE.** (a) Ratio of the AHE electric field $E_{AHE}^{3\omega}$ and the cubed longitudinal electric field $E_{xx}$ (left axis), and longitudinal conductivity $\sigma_{xx}$ (right axis) as a function of $T$. $E_{AHE}^{3\omega} = V_{AHE}^{3\omega}/L_{xy}$ with $L_{xy} \cong 21$ μm and $E_{xx} = V_{xx}/L_{xx}$ with $L_{xx} \cong 26$ μm. (b) $E_{AHE}^{3\omega}/(\sigma_{xx}E_{xx}^3)$ and fit to the data (solid line) plotted as a function of $\sigma_{xx}^2$. The respective temperatures associated with the data points are indicated.